# On the persistence of polar domains in ultrathin ferroelectric capacitors

**Pavlo Zubko[1*], Haidong Lu[2], Chung-Wung Bark[3], Xavi Martí[4], José Santiso[5], Chang-Beom Eom[3], Gustau Catalan[5,6*] and Alexei Gruverman[2*]**

[1] London Centre for Nanotechnology and Department of Physics and Astronomy, University College London, 17-19 Gordon Street, London, WC1H 0AH, UK
[2] Department of Physics and Astronomy, University of Nebraska-Lincoln, NE 68588, USA
[3] Department of Materials Science and Engineering, University of Wisconsin-Madison, WI 53706, USA
[4] Institute of Physics, Academy of Sciences of the Czech Republic, Cukrovarnická 10, 162 53 Praha 6, Czech Republic
[5] Institut Catala de Nanociencia i Nanotecnologia (ICN2), (CSIC-ICN), Barcelona Institute of Science and Technology, Campus Bellaterra, Barcelona 08193, Spain.
[6] Institut Catala de Recerca i Estudis Avançats (ICREA), Barcelona.

E-mails: agruverman2@unl.edu, gustau.catalan@icn2.cat and p.zubko@ucl.ac.uk

**Abstract.** The instability of ferroelectric ordering in ultra-thin films is one of the most important fundamental issues pertaining realization of a number of electronic devices with enhanced functionality, such as ferroelectric and multiferroic tunnel junctions or ferroelectric field effect transistors. In this paper, we investigate the polarization state of archetypal ultrathin (several nanometres) ferroelectric heterostructures: epitaxial single-crystalline $BaTiO_3$ films sandwiched between the most habitual perovskite electrodes, $SrRuO_3$, on top of the most used perovskite substrate, $SrTiO_3$. We use a combination of piezoresponse force microscopy, dielectric measurements and structural characterization to provide conclusive evidence for the ferroelectric nature of the relaxed polarization state in ultrathin $BaTiO_3$ capacitors. We show that even the high screening efficiency of $SrRuO_3$ electrodes is still insufficient to stabilize polarization in $SrRuO_3/BaTiO_3/SrRuO_3$ heterostructures at room temperature. We identify the key role of domain wall motion in determining the macroscopic electrical properties of ultrathin capacitors and discuss their dielectric response in the light of the recent interest in negative capacitance behaviour.

On the persistence of polar domains in ultrathin ferroelectric capacitors

**Preface**
This article is dedicated to Prof. Ekhard Salje, a pioneer in the field of domain wall physics and a constant source of inspiration to all of us. The first problem that one must face in the field of domain wall nanoelectronics, and one to which Ekhard himself has dedicated some effort [1,2,3,4], is defining when are domain walls stable, because in order to study the properties of domain walls, one must have stable domain walls in the first place. This question is typically formulated as: when does a material transition from a mono-domain state to a polydomain state?

The answer of course has to do with the balance between the energy gained by generating domains (which reduce depolarizing fields in ferroelectrics, or elastic stress in ferroelastics) and the energy cost of domain walls. The problem is that this balance of energies depends completely on boundary conditions, and these are often ill-determined; not so much for ferroelastics (as Ekhard likes to remind everybody, strain cannot be screened), but very much so for ferroelectrics, where free charges can screen the depolarizing field and thus the driving force for the appearance of domains. Ekhard in his wisdom realized this quickly, and sometimes confides that is the reason he has chosen work with ferroelastics instead of ferroelectrics, i.e., to make his life a bit easier… us ferroelectricians are not so fortunate and must deal with the problem of screening at the boundary.

The following paper is testimony to the complexity of this problem. It is the fruit of several years of research, which along the way led to the experimental demonstration of the tunnelling electroresistance effect in ferroelectrics [5] and discovery of polarization reversal via flexoelectric effect [6]. Numerous discussions of the authors among themselves and with colleagues in the community revealed that, surprisingly, there was not yet a clear consensus about even such basic questions as what is the dielectric constant that one should use when calculating the depolarizing field. We hope that our efforts will help clarify rather than muddy the waters.

**1. Introduction**
The last decade has seen a considerable transformation in our understanding of the fundamental limits of ferroelectricity at the nanoscale. In particular, it is now well understood that ferroelectricity can be stabilized in thin films that are only a few unit cells thick [7, 8, 9, 10]. This has opened up new device possibilities, such as exploiting the electric tunnelling that is possible at such reduced thicknesses. At the same time, though, the results emphasize the importance of boundary conditions: for films that are only a few unit cells thick, the boundary is a considerable percentage of the actual volume of the sample and has a very large impact on the functional behaviour. Imperfect screening of the depolarization field that arises at the surface of a ferroelectric can result in the formation of highly dense domain structures that can dominate the functional properties of the material [11, 12]. Understanding and controlling the polarization state of ultrathin films is not only of academic interest, but also essential for practical devices. While applications such as ferroelectric tunnel junctions require the stabilization of a single-domain polar state in ultrathin ferroelectric films sandwiched between electrodes [13], other devices may benefit from the formation of domains, which are known to greatly enhance the dielectric response of these materials [14]. Perhaps even more exciting is the prospect of harnessing the potential for applications of the many unexpected functionalities being discovered at domain boundaries [15, 16]. The recently emerged and rapidly developing field of domain boundary engineering, based on the idea that domain walls can host properties very different from the bulk and themselves act as the functional elements of new electronic devices, owes much to the work of Ekhard Salje. The discoveries of superconductivity at domain walls in $WO_{3-x}$ [17], ferrielectricity at domain walls in $CaTiO_3$ [18,19], fast ion transport along twin boundaries in $WO_3$ [20], and complex polar states at domain walls in $SrTiO_3$ [21,22] are just a few of his pioneering

On the persistence of polar domains in ultrathin ferroelectric capacitors contributions that have stimulated this new, exciting field of research. In this context, the ultradense domain structures that naturally appear in nanoscale ferroelectric films seem like an ideal route for engineering functional materials dominated by the properties of domain walls.

In this paper, we have studied the polarization state of an archetype for ultrathin capacitor structures [7,23,24,25]: epitaxial single-crystalline films of ferroelectric $BaTiO_3$ sandwiched between the most habitual perovskite electrodes, $SrRuO_3$, on top of the standard perovskite substrate, $SrTiO_3$ [26, 27]. The thickness of the $BaTiO_3$ films is only ~5 nm (12 unit cells) so as to prevent strain relaxation via misfit dislocations, and also to stay within a thickness range relevant for tunnelling devices. We have found that, though the ultrathin $BaTiO_3$ films behave functionally as paraelectric [28], they are in a ferroelectric phase with a strongly enhanced tetragonality. This is reconciled by the inferred presence of ferroelectric nanodomains that average the total polarization to zero, even though within each domain the polarization is large. The domain walls between these nanodomains are extremely mobile, reducing the coercivity to values below the modulation ac voltage used in dielectric and piezoelectric measurements. In this highly susceptible state, any small external input, such as a dc voltage or a tip-induced strain gradient, can lead to strong poling of the film. Indeed, decomposing the dielectric response into the individual contributions from the ferroelectric and the interface layers, suggests that the dielectric constant of the ferroelectric itself is actually negative over a significant range of temperatures. It is moreover found that, at cryogenic temperatures, the mobility of the domain walls is drastically reduced, so that a frozen bulk-like ferroelectric behaviour can be recovered. Our results corroborate previous findings of large enhancements in dielectric response due to nanoscale domain wall motion and at the same time suggest that where a single-domain state is required, strategies for stabilizing it should focus on reducing the mobility of the domain walls, for example by increasing the density of pinning centres.

## 2. Results and discussion

Given that the $BaTiO_3$ films used in the present study are fully coherent with the $SrTiO_3$ substrates, which impose compressive stress and thus enhance vertical tetragonality, it is expected that the films should be in the ferroelectric state with out-of-plane polarization, with only 180º antiparallel domains allowed [29]. Indeed, as-grown films without a top electrode are experimentally found to be ferroelectric, with the virgin state being single-domain with polarization pointing upward (i.e. away from the substrate). Figures 1a and 1b show a concentric square-shape domain pattern produced by scanning the virgin $BaTiO_3$ surface, first, with a tip under positive +3 V dc bias (large bright square) and, then, under negative dc bias of –3V (small dark square). Local PFM spectroscopy studies reveal a hysteretic switching behaviour typical for ferroelectric polarization reversal (figure 1c). The hysteresis loop is characterized by a slight shift toward a positive voltage suggesting the presence of a built-in electric field of about $10^6$ V/cm, which is responsible for the upward orientation (toward the free surface) of the polarization in the virgin films. It has been found that the PFM amplitude and phase images of the domain pattern in figures 1a and 1b showed almost no relaxation 3 days after poling, suggesting a highly stable polarization.

By contrast, quite a different behaviour is observed in the same $BaTiO_3$ films when they have 2 nm-thick top $SrRuO_3$ electrodes—the structures referred to as ultrathin $BaTiO_3$ capacitors. PFM imaging of the capacitor with a superimposed dc bias of $\pm 1$ V revealed a saturated amplitude signal, and phase contrast inversion accompanying the change in the dc bias polarity (figures 2a and 2b). Meanwhile, in the absence of bias, the same $BaTiO_3$ capacitor exhibits noise-level amplitude and phase signals (figures 2c and 2d), indicating that the net

On the persistence of polar domains in ultrathin ferroelectric capacitors polarization in the capacitors is close to zero when there is no external bias - in agreement with the earlier PFM imaging and polarization hysteresis loop measurements [28].

This behaviour indicates that the BaTiO$_3$ capacitor experiences strong polarization relaxation as a result of imperfect screening of the depolarizing field by SrRuO$_3$ electrodes. This also suggests that the depolarizing fields are stronger in the ultrathin capacitors than in the films with exposed top surfaces. That depolarization should be stronger in a capacitor than in the film with a bare surface may seem counter-intuitive: after all, it is expected that the charges on the electrodes would provide more effective screening of the depolarization field than the adsorbates on the exposed BaTiO$_3$ surface. On the other hand, it has previously been shown that adsorbates on free ferroelectric surfaces are in fact extremely effective in screening of polarization [30, 31] and can even allow chemically-induced switching [32, 33], while electrodes are never perfect in this respect due to their finite effective screening length [7, 34].

It is not clear, however, whether the depolarizing field in the capacitors causes a transition of BaTiO$_3$ into a true paraelectric phase, or a formation of small (less than the spatial resolution limit of PFM) antiparallel 180º domains. Note that PFM by itself cannot differentiate between these two scenarios: in either case, the zero-bias piezoelectric response would be suppressed as observed. In order to resolve this, we resort to structural characterization and electrical measurements of the SrRuO$_3$/BaTiO$_3$/SrRuO$_3$ capacitors. An X-ray diffraction off-specular reciprocal space map in figure 3a shows that the diffraction peak corresponding to BaTiO$_3$ is fully in-plane aligned with that of the SrTiO$_3$ substrate and the SrRuO$_3$ electrode. The peak position along $Q_x$ is the same for films and substrate, indicating that the epitaxial layers are coherent and dislocation-free [35]. This is important because it implies that the considerable in-plane compression imposed by the substrate is unrelaxed. This strain should enhance the tetragonality and thus also the ferroelectricity of the BaTiO$_3$ layer. Closer inspection in the form of θ–2θ scans (figure 3b) yields further information in this respect. A fit of the X-ray diffraction peaks in figure 3b using dynamical theory yields out-of-plane lattice parameters of 4.141±0.004 Å and 3.953±0.002 Å for BaTiO$_3$ and SrRuO$_3$, respectively. Note that the fitting also refines the layer thicknesses, which are in excellent agreement with X-ray reflectivity data (not shown). We emphasize the need for a fitting of the complete layered heterostructure, which cannot be circumvented by a more straightforward pick of the intensity maxima around the expected BaTiO$_3$ position [36, 37]; this erroneous procedure would lead to ~4.08 Å (~0.5° off in 2θ), which is much smaller than the actual value, and would lead to a different and incorrect interpretation of the results.

The out-of-plane lattice parameter of the BaTiO$_3$ film is too big to be compatible with a paraelectric phase under in-plane strain. An extrapolation of the bulk paraelectric lattice parameter to room temperature yields a pseudocubic value of $a_0$=4.006 Å [38]; using the elastic constants of BaTiO$_3$ from Ref. 39, the Poisson's ratio expansion of the paraelectric unit cell caused by the in-plane compression should lead to an out-of-plane lattice parameter of only $c_0$ = 4.102 Å, which is more than 1% smaller than the experimentally measured value $c$=4.141 Å. The extra tetragonality is consistent with the BaTiO$_3$ being ferroelectric. The additional strain $e_{33}=(c-c_0)/c_0=9.5\times10^{-3}$ can be related to the polarization by $P^2=e_{33}/Q_{11}$ [40, 41], where $Q_{11}$ is the electrostrictive coefficient. Using the electrostrictive coefficient for BaTiO$_3$ ($Q_{11}$= 0.11 m$^4$C$^{-2}$) [39], the measured out-of-plane strain corresponds to a polarization of 29 μC/cm$^2$, comparable to the polarization of bulk BaTiO$_3$ at room temperature (~26 μC/cm$^2$) [42]. Naturally, such an estimate is only as accurate as the materials parameters used in the calculation, which are known to vary somewhat from sample to sample (e.g. see Ref. 29). However, our tetragonality is also in good agreement with values observed experimentally by Petraru *et al.* [43] on similar SrRuO$_3$/BaTiO$_3$/

On the persistence of polar domains in ultrathin ferroelectric capacitors

SrRuO$_3$ capacitors. For structures that were fully strained to the SrTiO$_3$ substrate, they measured out-of-plane lattice parameters in the range 4.14–4.17 Å, and spontaneous polarization values of 35–45 μC/cm$^2$. On the other hand, earlier studies by Kim *et al.* [24] and Yanase *et al.* [44] recorded much larger lattice parameters ($c \approx$ 4.24–4.27 Å and $c \approx$ 4.35–4.37 Å respectively) for their samples with similar polarization values.

Since defects (e.g. oxygen vacancies) can also cause an expansion of the lattice, to verify that indeed our films are ferroelectric at room temperature we turn to functional (dielectric and ferroelectric) measurements. First, the dielectric constant (figure 4a) decreases with decreasing temperature, indicating that the dielectric maximum—a signature of a ferroelectric phase transition—is above room temperature. Second, figure 4b shows polarization hysteresis loops for the sample at room temperature and at 4.2 K. The room-temperature polarization loop is only weakly hysteretic with negligible remnant polarization and "coercive voltages" of less than 0.1 V. At this temperature the domain walls are very mobile, with a bias of only 0.5 V being sufficient to almost fully saturate the polarization. Extrapolation of the saturation polarization to zero bias gives an estimate for the spontaneous polarization of 31-33 μC/cm$^2$, in excellent agreement with the value calculated from the film's tetragonality. Notice also that application of the imaging PFM bias of only 0.2 V (peak-to-peak) would make the domain walls oscillate near equilibrium positions, rendering their visualization impossible even if the domain size were comparable to the resolution limit of PFM.

Invoking domain wall mobility is also important for understanding the dielectric behaviour. Nanoscale domain wall motion gives rise to large extrinsic contributions to the dielectric permittivity (figure 4a) as was previously reported for polydomain PbTiO$_3$/SrTiO$_3$ superlattices [45]. At low temperature, pinning of the domain walls reduces their contribution to the dielectric response [46], causing it to decrease to values closer to the intrinsic dipolar response (which, from Landau theory calculations, should be around $\varepsilon_r \approx 20$ at 4 K for epitaxial BaTiO$_3$ on SrTiO$_3$). The low temperature pinning of the walls also opens up the ferroelectric hysteresis loop: the coercive voltage is about 1.5V (the equivalent coercive field is 3 MV/cm), more than an order of magnitude larger than at room temperature. The remnant polarization at 4.2 K is ~30 μC/cm$^2$, in remarkable agreement with the 0 K theoretical prediction of 31 μC/cm$^2$ [7]. In summary, then, the functional characterization fully supports the view that the films are ferroelectric, but with highly mobile domain walls at room temperature that, combined with the depolarizing fields, cause a relaxation of the polarization.

The very high resistance to dielectric breakdown of the films, without which the films would suffer dielectric rupture before switching, is also noteworthy: the highest applied voltages in this work correspond to the electric fields approaching the breakdown strength of high-quality single crystals [47]. Just as remarkable, the high value (3 MV/cm) of the experimentally measured coercive field at cryogenic temperatures is of the order of the thermodynamic coercive field obtained from standard Landau theory for intrinsic switching. Using Landau theory, the coercive field of BaTiO$_3$ under compressive strain from SrTiO$_3$ is calculated to be 1.88 MV/cm at room temperature and 3.75 MV/cm at 4 K. However, these values should be treated with great caution as the actual voltage across the ferroelectric layer will be very different due to the voltage drop across the interfacial layers [48].

It is also useful to consider how large the domains in the BaTiO$_3$ capacitors may be. In the absence of screening, the domain size $w$ would be given by the universal expression [49,16]

On the persistence of polar domains in ultrathin ferroelectric capacitors

$$w^2 = \frac{\sqrt{2}\pi^3}{21\zeta(3)} \sqrt{\frac{\varepsilon_\perp}{\varepsilon_\parallel}} d\delta, \qquad (3)$$

where $\delta$ is the domain wall half-width, $\varepsilon_\parallel$ and $\varepsilon_\perp$ are the dielectric constants parallel and perpendicular to the polarization direction, and $\zeta(3) = 1.18$. Using the $\delta$ value of ~4 Å [50] and $\sqrt{\varepsilon_\perp/\varepsilon_\parallel} \approx 5$ yields an equilibrium domain size of around 30 Å. Experimentally, for a 10 nm-thick BaTiO$_3$ film sandwiched between SrTiO$_3$ layers, Tenne *et al.* [51] observe a domain periodicity of 63 Å. Thus, for a 5 nm film, a Kittel-like square root extrapolation of their result yields a domain period of $63/\sqrt{2}$ Å = 44.5 Å (i.e. a domain size of ~22 Å), in reasonable agreement with the theoretical estimate from equation 3. On the other hand the presence of SrRuO$_3$ electrodes in our samples should in principle reduce the depolarizing field, so one might expect the domain size to be much larger than in the case of unscreened capacitors considered above. Yet a square root extrapolation based on the *ab-initio* results of Aguado-Puente and Junquera [52], who simulated the domain structure in SrRuO$_3$/BaTiO$_3$/SrRuO$_3$ capacitors, also leads to $w$ in the range of 20-60 Å. Thus, all evidence suggests that it is reasonable to expect the domain size in our BaTiO$_3$ capacitors to be of the order of few nanometers and well below the resolution limit of PFM.

The stability of the polydomain state implies that the screening of the spontaneous polarization of BaTiO$_3$ by free charges from SrRuO$_3$ is incomplete. This is not surprising as, even for structurally perfect materials, the screening length is finite [7,34], and a transition to polydomain structures in ultrathin capacitors has in fact been reported already by Nagarajan *et al.* for the perovskite ferroelectric, Pb(Zr,Ti)O$_3$ [12]. Let us therefore consider the magnitude of the depolarizing field $E_{dep}$ in the system and thus get a rough estimate for the degree of screening provided by the SrRuO$_3$ electrodes.

An upper limit for $E_{dep}$ can be obtained from the value of the remnant polarization $P_r$ at 4.2 K. Suppose $E_{dep} = \alpha P_r/\varepsilon_0$, where $\alpha = 0$ in the case of perfect screening and $\alpha = 1$ for open-circuit boundary conditions. This depolarizing field cannot exceed the intrinsic coercive field - otherwise the remnant polarization would be unstable; therefore $E_{dep} < E_{int(4\ K)} = 3.75$ MV/cm. Combining these two relationships, we estimate the upper bound for the screening factor $\alpha$:

$$\alpha \lesssim E_{int} \frac{\varepsilon_0}{P_r} \qquad (4)$$

At 4.2 K, $P_r \approx 30$ µC/cm$^2$ gives $\alpha \lesssim 0.01$, i.e. the SrRuO$_3$ electrodes are more than 99% efficient at screening the polarization. By comparison, a recent theoretical study of 7-nm-thick (rhombohedral) BaTiO$_3$ films has found that the striped domains give way to the monodomain state for screening efficiencies above 98% [53].

The parameter $\alpha$ can alternatively be expressed in terms of the effective screening length $\lambda$ in the electrodes as $\alpha = 2\lambda/d_{BTO}$ [54], giving an upper limit for $\lambda$ of 27 pm. This value can be directly compared with several theoretical predictions. For instance, frozen ion DFT calculations for SrRuO$_3$/BaTiO$_3$/SrRuO$_3$ capacitors give $\lambda = 24$ pm [7], whereas a full relaxation reduces this value to ~10 pm [55]. Experimentally, a value of 12 pm was obtained for the PbTiO$_3$/Nb-doped SrTiO$_3$ interfaces by fitting the dependence of lattice parameters on film thickness [56]. Our estimate is comparable with all these values, suggesting a high quality of the SrRuO$_3$ electrodes.

The finite screening length has another peculiar effect on the dielectric response. The interface acts a capacitance $C_i$ in series with that of the BaTiO$_3$ layer ($C_{BTO}$) and it is instructive to try to separate out these two contributions to the overall measured response.

On the persistence of polar domains in ultrathin ferroelectric capacitors

The measured inverse capacitance per unit area is $\frac{A}{C} = 2\frac{A}{C_i} + \frac{A}{C_{BTO}}$. For an estimate of the interface contribution we take the value obtained by first principles calculations of Stengel et al. $\frac{A}{C_i} = 2.28$ m$^2$/F [57], which compares reasonably well with above-room-temperature values obtained experimentally for SrRuO$_3$/Ba$_{0.7}$Sr$_{0.3}$TiO$_3$/SrRuO$_3$ capacitors [58]. Our measured $\frac{A}{C}$ values are 2.18 m$^2$/F at 300 K and 34 m$^2$/F at 5 K. Neglecting the possible temperature dependence of $C_i$, we can estimate the $\frac{A}{C_{BTO}}$ values as 30 m$^2$/F and $-2.4$ m$^2$/F at 5 K and 300 K respectively. At low temperature, the interface-corrected capacitance of BaTiO$_3$ is almost the same as the measured value. At room temperature, however, the BaTiO$_3$ capacitance obtained from our calculation turns out to be negative. Although this result at first appears unphysical, it is precisely what is expected for a ferroelectric with depolarization-field-induced domain structure, as first discussed by Bratkovsky and Levanyuk [59] and more recently in Refs 60 and 14 (and by Stengel et al. [57] for the monodomain case). Again, with the assumption of a temperature independent $C_i$, the cross-over from positive to negative capacitance in our BaTiO$_3$ layer would occur around 160 K. While the capacitance of the system as a whole is positive, as required for thermodynamic stability, it has been suggested that the local enhancement of the potential at the metal-ferroelectric (or the analogous semiconductor-ferroelectric) interface due to this negative capacitance effect may be very useful in reducing the power consumption of field effect transistors [61].

It is worth emphasising that the above analysis relies on a theoretical value for the interface capacitance and the calculated capacitance for the BaTiO$_3$ layer would become positive if the interface capacitance were higher by more than a factor of 2. We note, however, that the slope of the polarization-voltage curve for our 12 unit cell thick BaTiO$_3$ capacitor is very similar to that of the 24 unit cell thick capacitor reported in Ref. 28 fabricated under the same conditions. This demonstrates that the measured capacitance of these samples is dominated by that of the interfaces and that the capacitance of the BaTiO$_3$ layers, if not negative, is at the very least much larger than that of the interfaces.

These results highlight the fact that, even when the screening efficiency of the metal-ferroelectric interface is very high and the screening length is very short, the residual depolarizing field is still sufficient to destabilize the polarization at room temperature and cause the formation of domains. In the presence of intrinsically incomplete screening, then, efficient domain wall pinning is required to ensure the stability of a poled state, a result highlighted by the stability of the remnant polarization at low temperatures where the walls are frozen. Variations in the pinning strength can lead to dramatic differences in functional properties and could help reconcile our results with those of Kim et al. [24, 25], who observe a similar saturation polarization but significantly larger remnant polarization and coercive field values in their 5 nm-thick SrRuO$_3$/BaTiO$_3$/SrRuO$_3$ capacitors, possibly due to more effective pinning of the domain walls (consistent with their slow relaxation dynamics). This also suggests that the stability of the poled state in films without a top electrode may perhaps be due to increased pinning of domain walls at surfaces, rather than (or as well as) to more efficient screening by atmospheric adsorbates.

While our results demonstrate that the stability of the poled state at low temperature is accompanied by a reduction in domain wall mobility, it may also be correlated to a simultaneous suppression of reverse domain nucleation. Further investigation is required to decouple the effects of nucleation and domain wall pinning.

On the persistence of polar domains in ultrathin ferroelectric capacitors

## 3. Summary and conclusions

Our findings corroborate that stabilizing ferroelectricity in ultrathin capacitors at room temperature is *not* fundamentally difficult, and can be achieved using epitaxial strain, as shown here and in previous works [7, 24, 25]. However, stabilizing a saturated polar state *is* difficult due to the inevitable presence of depolarizing fields: as we have noted, even electrodes with screening efficiencies around 98-99% may not be efficient enough. Combined with the high mobility of the domain walls in good quality samples with few pinning defects, the result is the breakdown of the poled state into antiparallel nanodomains that average the polarization to zero. This implies that in order to obtain a robust polar state we may perhaps need films with more (not less) defects, so that the domain walls are more effectively pinned. Suppression of reverse domain nucleation, however, may also play an important part in the stability of the poled state at low temperature and further study is required to elucidate its role.

While the polydomain state is undesirable for applications in tunnelling devices, the enhanced dielectric response demonstrated in this work and the possibility of negative capacitance behaviour is of direct interest for improving the power consumption of conventional field effect transistors.

The bottom line, going back to Ekhard's dictum, is that although, unlike strain, polarization can be screened, the screening can never be perfect, and even the tiniest fraction of depolarizing field is sufficient to cause the spontaneous appearance of domains. The problem of stabilizing a polar state is thus one of pinning the position of the domain walls rather than one of preventing their inevitable appearance.

## 4. Experimental

*Sample preparation*: Thin films of $BaTiO_3$ were fabricated by pulsed laser deposition assisted by reflective high energy electron diffraction (RHEED) in order to ensure layer-by-layer growth with atomically flat interfaces. All the $BaTiO_3$ thin films were coherently grown on $SrTiO_3$ substrates with a conductive buffer layer of $SrRuO_3$ (50 nm) that served as a bottom electrode. Prior to deposition, the $SrTiO_3$ substrates (miscut angle < 0.1º) were etched using buffered HF acid for 90 seconds and annealed in oxygen for 12 hours at 1000ºC to ensure atomically smooth, $TiO_2$-terminated surfaces. The oxide layers were grown at 680ºC in 150 mTorr of oxygen. Films both with exposed top surfaces and with deposited top $SrRuO_3$ electrodes have been used in this study in order to compare the effect of boundary conditions on polarization. The 2 nm-thick top $SrRuO_3$ layers were deposited on top of the $BaTiO_3$ film at 600ºC without breaking the vacuum cycle and patterned into pads of various sizes using a standard photo-lithography technique. After etching the photoresist away by Ar-ion milling the samples were rinsed in acetone and IPA and then dried with $N_2$ gas.

*Electrical and structural characterization:* Atomic force microscopy imaging of the films confirms the atomically flat surface of the samples with the unit-cell-high vicinal steps of the substrate being reproduced on the top surface of the film. Topographic and PFM measurements of the films were performed using an atomic force microscope MFP-3D (Asylum Research) equipped with conductive tips (DPE18/Pt, Mikromasch). PFM hysteresis loops were obtained at fixed locations on the sample surface as a function of dc switching pulses (25 ms) with a superimposed ac modulation bias of 1.0 $V_{p-p}$ at 300 kHz.

X-ray diffraction measurements were performed in a four-circle diffractometer (PANalytical X'Pert MRD) with $CuK_{\alpha 1}$ radiation (wavelength=1.5406 Å, with 4-crystal Ge(220) monochromator). Dielectric measurements were performed using an Agilent 4284A precision LCR meter with an ac driving bias of 10 mV at 1 kHz. Ferroelectric polarization-voltage

On the persistence of polar domains in ultrathin ferroelectric capacitors loops were measured with an aixACCT TF Analyzer 2000 ferroelectric tester using a 1 kHz triangular waveform. 60 µm-diameter electrodes were contacted using indium wire and the temperature was varied by slowly lowering the sample into liquid He.

The Landau theory coefficients and elastic constants for $BaTiO_3$ used in this work are $T_0 = 110°C$, $(2C\varepsilon_0)^{-1} = 3.3\times10^5$ $K^{-1}C^{-2}Nm^2$, $Q_{11} = 0.11$ $m^4C^{-2}$, $Q_{12} = -0.043$ $m^4C^{-2}$, $s_{11} = 8.3\times10^{-12}$ $m^2N^{-1}$, $s_{12} = -2.7\times10^{-12}$ $m^2N^{-1}$ [39].

**Acknowledgements**
H.L. and A.G. acknowledge support by the National Science Foundation (NSF) through Materials Research Science and Engineering Center (MRSEC) under Grant DMR-1420645. P.Z. would like to thank Prof. Jean-Marc Triscone for useful discussions and for making possible the electrical measurements reported in this work, and acknowledges funding from EPSRC (Grant No. EP/M007073/1). The work at University of Wisconsin-Madison (C.W.B and C.B.E.) was supported by the AFOSR under Grant FA9550-15-1-0334. G.C. acknowledges funding from project FIS2013-48668-C2-1-P. G.C. and J.S. acknowledge support from the Severo Ochoa programme. X.M. acknowledges support from the Grant Agency of the Czech Republic (Grant No. 14-37427). P.Z., G.C. and A.G. thank the Leverhulme Trust for international network funding (F/00 203/V) for the funds that have enabled this collaboration.

On the persistence of polar domains in ultrathin ferroelectric capacitors

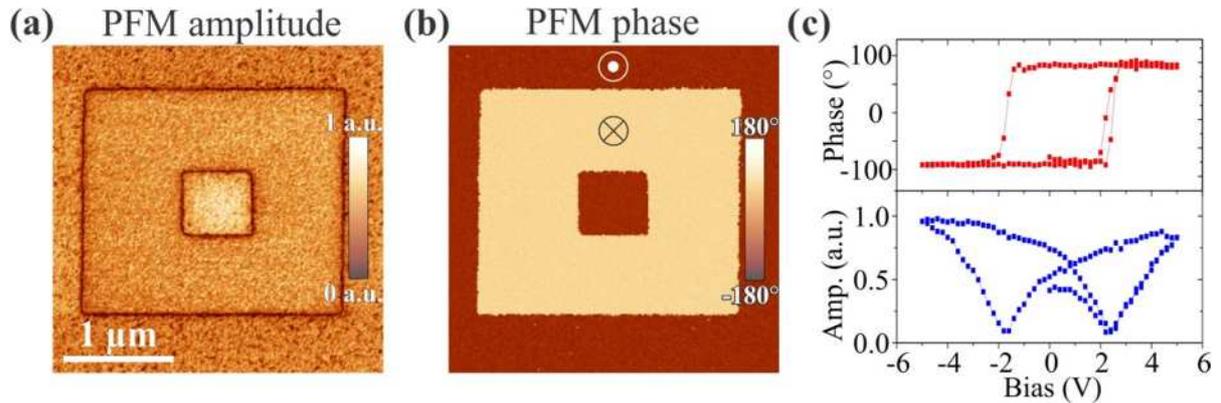

**Figure 1.** Ferroelectric switching in a 5-nm-thick BaTiO$_3$ film with a bare surface: (a) PFM amplitude and (b) phase signals for domains written with +3 V (large square) and –3 V (small square) dc bias. The unwritten background is homogeneously polarized upward. (c) Local PFM hysteresis loops measured for the same film, illustrating reversible switching of the ferroelectric polarization (top - phase signal, bottom - amplitude signal).

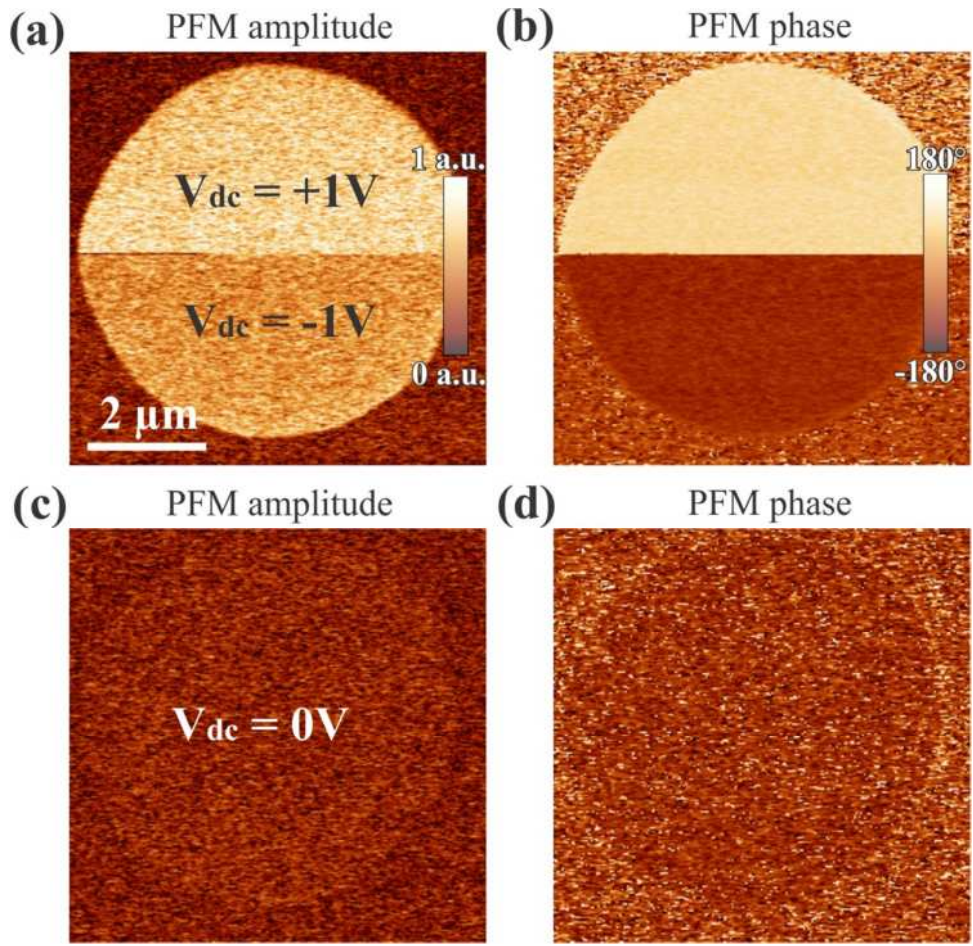

**Figure 2.** PFM amplitude (a, c) and phase (b, d) images of 5-nm-thick BaTiO$_3$ capacitor obtained while scanning over the circular SrRuO$_3$ top electrode with the tip under ±1 V dc (a, b) and 0 V dc (c, d) bias.

On the persistence of polar domains in ultrathin ferroelectric capacitors

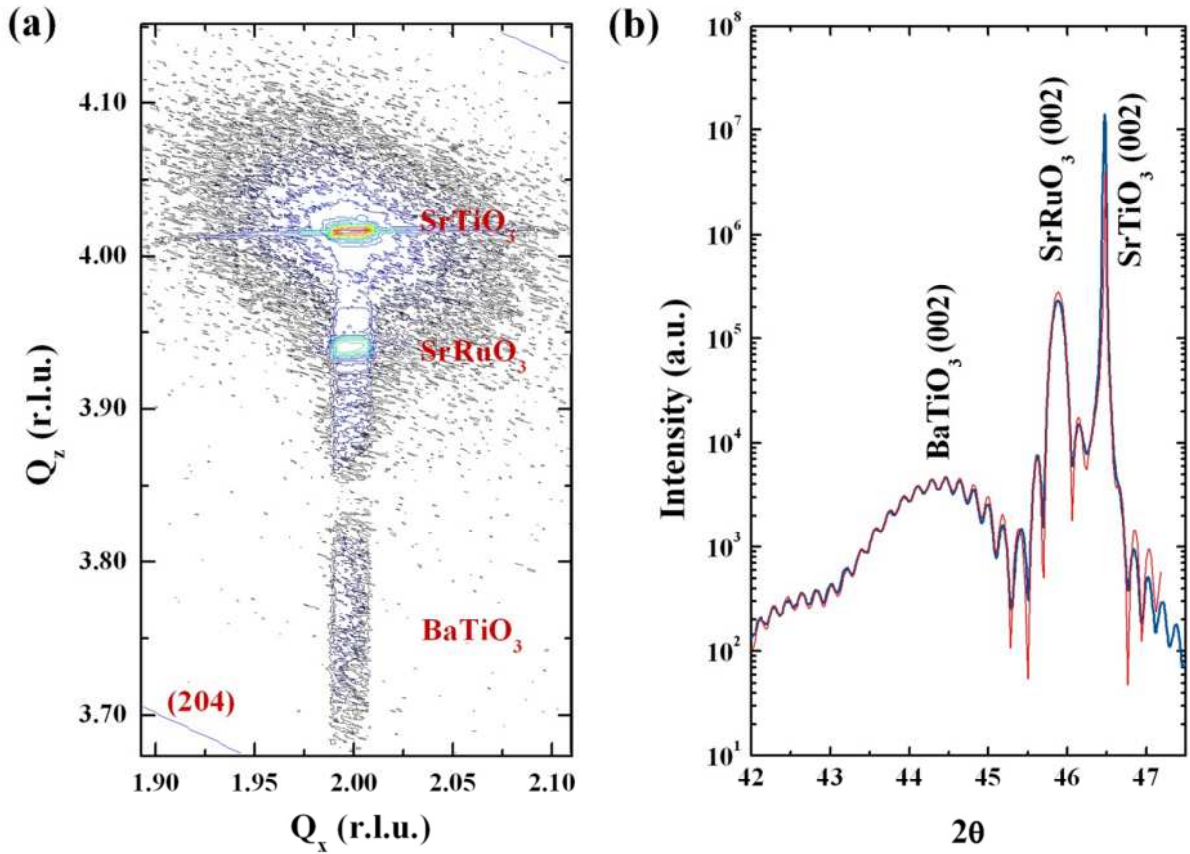

**Figure 3.** Results of X-ray diffraction characterization of the 5-nm-thick $SrRuO_3/BaTiO_3/SrRuO_3$ capacitors: (a) Reciprocal space map around the (204) reflections of the $SrTiO_3$ substrate, and $SrRuO_3$ and $BaTiO_3$ films confirming fully coherent growth. (b) Intensity profile along the specular crystal truncation rod around the (002) reflections of the $SrTiO_3$ substrate, the $SrRuO_3$ electrodes and the $BaTiO_3$ thin film.

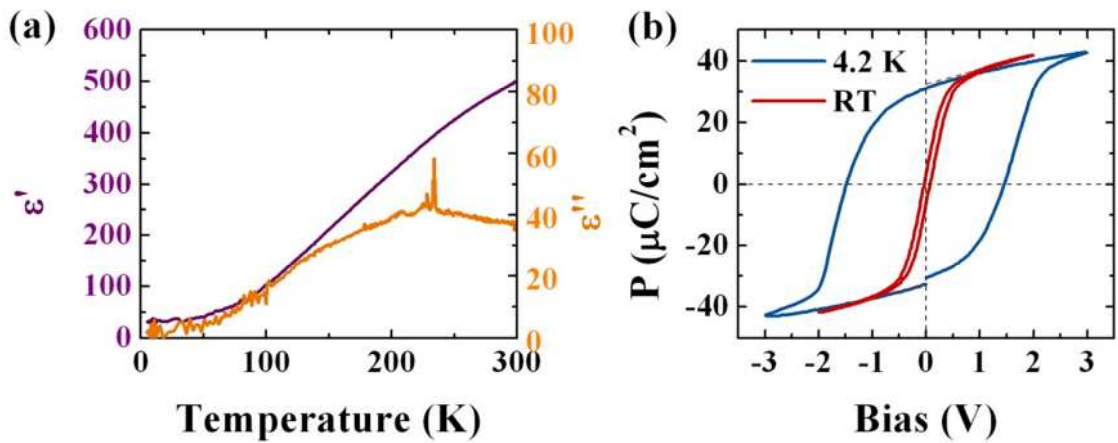

**Figure 4.** (a) Effective dielectric constant and loss obtained directly from measurements of the sample capacitance $C = \frac{\epsilon\epsilon_0 A}{d_{BTO}}$ as a function of temperature. (b) Polarization hysteresis loops at room temperature (RT) and at 4.2 K.

On the persistence of polar domains in ultrathin ferroelectric capacitors

On the persistence of polar domains in ultrathin ferroelectric capacitors